\documentclass[sigconf, 9pt]{acmart}

\usepackage{graphicx}
\usepackage{amsmath,amsfonts}
\usepackage{mathtools}
\usepackage{comment}
\usepackage[subrefformat=parens,labelformat=parens]{subfig}

\usepackage{caption}
\usepackage{subcaption}

\usepackage{bm}
\usepackage{multirow}
\usepackage{threeparttable,booktabs}
\usepackage{tikz}
\usepackage{balance}
\usepackage{courier}                                       
\usepackage{cleveref}                                      
\usepackage[mathcal]{eucal}

\usepackage{scalefnt}
\makeatletter\let\float@addtolists\relax\makeatother

\usepackage{algorithmic}
\usepackage{algorithm}

\usepackage{pgfplots}
\usepackage{pgfplotstable}
\pgfplotsset{compat=newest}
\usepackage{siunitx}
\theoremstyle{plain}

\theoremstyle{definition}

\usepackage[skip=1pt]{caption}            
\newrobustcmd{\B}{\bfseries}

\usepackage{tabularx}

\usepackage{multirow}
\usepackage{color}

\usepackage{array}
\newcolumntype{C}[1]{>{\centering\arraybackslash}p{#1}}  

\definecolor{hyunsu}{RGB}{210, 100, 0}

\definecolor{chris}{RGB}{128, 0, 0}

\definecolor{Houbo}{RGB}{128, 0, 128}

\newcommand{\workname}{FoundaRF}

\usepackage[table]{xcolor}
\usepackage{amsmath}

\usepackage{adjustbox}
\graphicspath{{./figs/}}          

\copyrightyear{2026}
\acmYear{2026}
\setcopyright{cc}
\setcctype{by-nc-nd}
\acmConference[ICCAD '26]{IEEE/ACM International Conference on Computer-Aided Design}{November 08--12, 2026}{San Jose, CA, USA}
\acmBooktitle{IEEE/ACM International Conference on Computer-Aided Design (ICCAD '26), November 08--12, 2026, San Jose, CA, USA}
\acmDOI{10.1145/3831252.3834018}
\acmISBN{979-8-4007-2873-0/2026/11}

\newcommand\blfootnote[1]{%
  \begingroup\renewcommand\thefootnote{}\footnote{#1}\addtocounter{footnote}{-1}\endgroup}

\title{FoundaRF: A Physics-Informed Foundation Model Framework for Radio-Frequency Passive Components}

\author{%
  Houbo~He\textsuperscript{1*\dag}, Chris~Han\textsuperscript{1*}, Hyunsu~Chae\textsuperscript{2*},
  Lei~Xia\textsuperscript{1}, Qiang~Zhou\textsuperscript{1}, Lawrence~Liu\textsuperscript{1},
  Sensen~Li\textsuperscript{2}, David~Z.~Pan\textsuperscript{2}, Taiyun~Chi\textsuperscript{1}}
\affiliation[obeypunctuation=true]{%
  \institution{\textsuperscript{1}Rice University, Houston, TX, USA\quad
               \textsuperscript{2}The University of Texas at Austin, Austin, TX,} \country{USA}}
\renewcommand{\shortauthors}{He et al.}

\begin{document}





\begin{abstract}
Designing RF/mmWave passive components is a time-consuming process that relies heavily on expert knowledge and iterative EM simulations. 
Although recent ML surrogate models have shown great promise in accelerating passive design, they remain task-specific and lack generalization. 
This paper presents~\workname{}, a foundation model framework that unifies passive layout modeling across diverse layout structures and technologies. 
It combines a self-supervised encoder with adaptive layout processing and physics-aware decoders to enable few-shot transfer. 
Experiments across four semiconductor technologies show strong generalization to unseen structures, outperforming baselines with up to a 3.3$\times$ reduction in prediction error, reducing labeled-data requirements by up to fourfold, and substantially shortening design turnaround. \workname{} establishes a scalable and data-efficient solution for RF/mmWave passive design automation.

 \vspace{-1em}

\end{abstract}

\maketitle
\blfootnote{\textsuperscript{*}These authors contributed equally to this work.
\newline
\textsuperscript{\dag}Corresponding author: Houbo He (houbo.he@rice.edu)}


\vspace{-3em}

\section{Introduction} \label{sec:intro}

Radio-frequency (RF) and millimeter-wave (mmWave) integrated circuits (ICs) are the backbone of today's wireless/wireline communication links and radar/sensing systems. Unlike low-frequency analog ICs, RF and mmWave IC performance is not governed by transistor characteristics alone; instead, it is often dominated by passive components, which perform key functionalities such as impedance matching, filtering, and signal conditioning \cite{pozar,razavi,MWSCAS,XiaohanJSSC2024,rfagent,haoisscc2025,HuLNA}. 
Designing these passive components, however, remains a critical bottleneck in the RFIC design flow, as it involves slow, iterative electromagnetic (EM) simulations, along with substantial designer expertise.

Recent advances in machine-learning~(ML)-based surrogate models have shown great promise in accelerating the design of passive components \cite{RFEDA_ASPDAC25_Chae,KaushikNC,chenhao_review}. However, these efforts typically focus on narrow, specific component families (e.g., transformers \cite{RFEDA_MWTL25_Chai,MOTIF-RF,yaolongIMS,hua2020,chenhao_ims2025}, inductors \cite{PACOSYT,Yaolong_DAC,shikai_iccad}, or resonators and filters \cite{resonator1,resonator2,shibata2022novel,jingyun_filter,dina_filter}), each tied to a specific process technology (e.g., CMOS or InP) and layout encoding (e.g., tabular design parameters~\cite{hua2020,PACOSYT,Yaolong_DAC,resonator2}, layout images~\cite{RFEDA_MWTL25_Chai,shibata2022novel,jingyun_filter}, or graphs~\cite{dina_filter}).
As a result, introducing a new passive topology or migrating to a different process technology requires building a new custom model from scratch, significantly limiting the scalability and generalizability of existing models.

Foundation models have transformed fields such as natural language processing and computer vision, where a single pretrained model learns transferable representations that can be adapted to a wide range of downstream tasks.
RF/mmWave IC design faces a similar need: \textbf{a unified model capable of operating effectively across various component families, layout topologies, frequency bands, process technologies, and physical dimensions.}
Despite the wide diversity of passive components, they share two universal properties: (1) their circuit behaviors are characterized by S-parameters, and (2) their underlying physics are governed by the same Maxwell’s equations.
These shared foundations suggest the feasibility of training a general-purpose model to capture common EM behaviors across various passive structures.

This paper presents \workname, a foundation model framework that, for the first time, generalizes RF/mmWave passive surrogate modeling across topologies, physical dimensions, frequencies, and semiconductor technologies. 
This foundation model framework (1) significantly reduces the data requirement for new tasks, (2) eliminates the need to share proprietary datasets, and (3) enables a scalable path toward RF/mmWave design automation.
From RF/mmWave IC designers' perspective, FoundaRF allows rapid performance evaluation of their designs and can be readily integrated into optimization loops for automated RF layout synthesis.

The major contributions of this work include:
\begin{itemize}

    \item \textbf{Unified foundation model framework.}
    \workname~employs a universal layout encoder and decoder paired with lightweight, task-specific adapters to extract broadband S-parameters across diverse RF/mmWave passive structures.

    \item \textbf{Adaptive, topology-focused representation.}
    An adaptive layout processing pipeline with distance field (DF)-based self-supervision is proposed, enabling the encoder to extract scale-invariant and template-agnostic topological features.

    \item \textbf{Physics-aware conditioning.}
    A Feature-wise Linear Modulation (FiLM)-conditioned decoder incorporates scaling and template metadata through channel-wise affine modulation, improving cross-domain robustness.

    \item \textbf{Strong cross-domain transfer.}
    Experiments demonstrate reliable performance on InP and SOI datasets and effective few-shot adaptation to SiGe and GaN, outperforming training-from-scratch and encoder-only transfer baselines.

    \item \textbf{Inverse design with rapid turnaround.} FoundaRF is demonstrated on two real-world T-coil synthesis tasks, a driver output network and an RX input ESD protection network, delivering EM-verified designs over 100$\times$ faster than EM-in-the-loop optimization while outperforming scratch-trained baselines in design quality.
\end{itemize}

\section{Background} \label{sec:prelim}

In this section, we provide background on RF passive design~(Sec.~\ref{sec:prelim:rf_passives}) and review the foundation model paradigm (Sec.~\ref{sec:prelim:found_models}).
We then present our datasets, including newly generated data to support generalization across technologies and structures~(Sec.~\ref{sec:prelim:dataset_generation}).

\vspace{-0.5em}
\subsection{Complexity in RF Passive Design} \label{sec:prelim:rf_passives}
RF/mmWave passive components typically consist of lumped elements (such as inductors, transformers, and capacitors) and distributed elements (such as transmission lines and couplers) \cite{pozar,razavi}.
Their physical layouts vary widely, depending on RF performance targets, semiconductor technologies, metal stacks, operating frequencies, and even designer preferences.
This results in a fragmented design landscape, where even similar functional blocks may look and behave quite differently across contexts. A concrete illustration of this variability can be seen in \cite{chenhao_ims2025}, demonstrating that even with the same template seed, a matching network can manifest dramatically different S-parameters depending on pixel filling rate and layer-to-layer interactions.

This variability extends beyond just the layout. In existing ML surrogate models, passive components are encoded in many forms, such as images, tabular parameters, graphs, or pixelated representations. Such fragmentation makes it impractical to build separate surrogate models for each combination of topology, technology, and layout encoding. A foundation model is therefore essential to enable generalizability.

\vspace{-0.5em}
\subsection{Foundation Models} \label{sec:prelim:found_models}

\begin{figure}[t]
\centering
\includegraphics[width=1\linewidth,height =4.3cm]{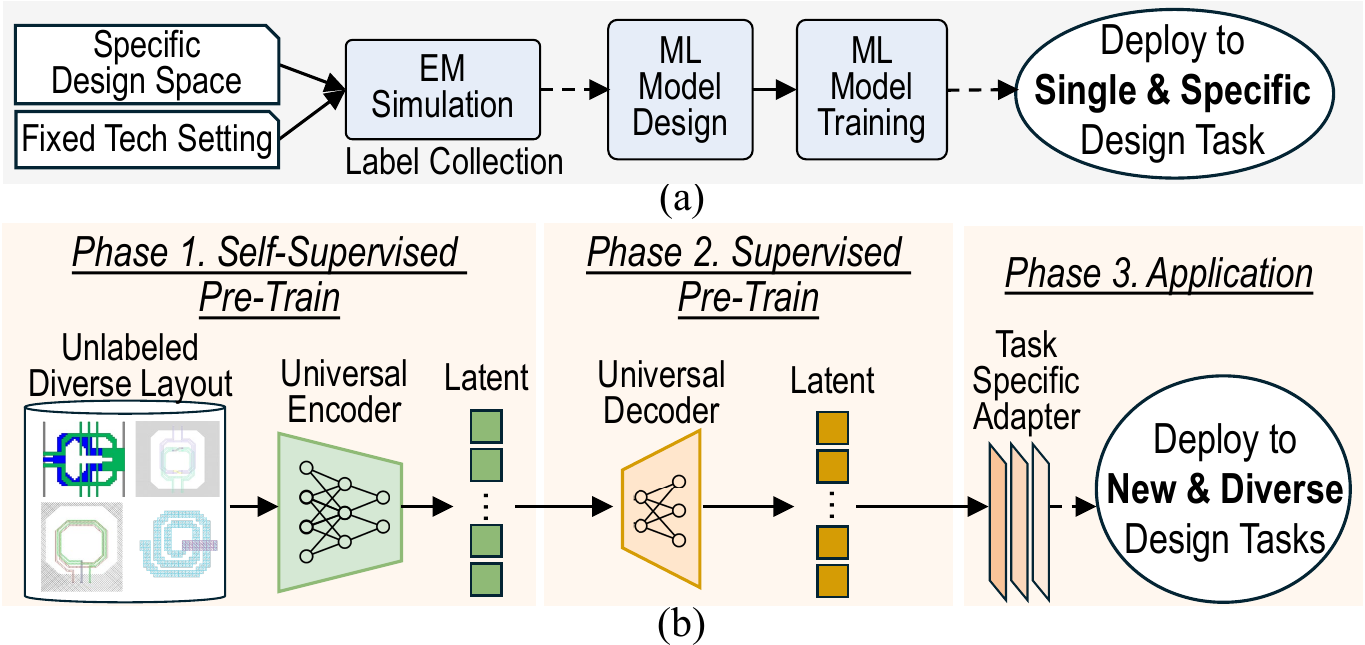}
\caption{Different ML modeling  paradigms: (a) task-specific fully supervised vs. (b) proposed FoundaRF framework.}
\label{fig:intro}
\vspace{-0.5em}
\end{figure}

Foundation models are large-scale neural networks trained on broad, diverse datasets to learn general-purpose representations that can be adapted to a wide range of specific or specialized tasks.
This approach has enabled breakthroughs in natural language processing~\cite{NEURIPS2020_gpt3}, computer vision~\cite{vision_foundation}, and EDA~\cite{fang2025surveycircuitfoundationmodel,progsg,liu2025deeprtl, fang2025circuitfusion,fang2025encoder}. 
These models benefit from self-supervised representation learning across large, unlabeled datasets, significantly reducing the need for hand-crafted, task-specific architectures or extensive supervision.

A similar approach can be applied to build a foundation model for RF passive components. 
However, unlike digital circuits, where functional equivalence, logic transformations, and cross-stage consistency (e.g., RTL $\leftrightarrow$ netlist) naturally provide self-supervised signals \cite{fang2025surveycircuitfoundationmodel,liu2025deeprtl,fang2025circuitfusion}, an RF foundation model would require \textbf{physics-aware supervision} that connects layout geometry to its frequency-dependent EM behavior. 
To address this, we adopt a hybrid training strategy that combines self-supervised pretraining and supervised pretraining. In contrast to prior task-specific supervised modeling (Fig.~\ref{fig:intro}(a)), our approach follows a three-phase workflow (Fig.~\ref{fig:intro}(b)):
(1) pretraining a universal encoder on unlabeled, diverse layout datasets to capture topology-aware representations, (2) pretraining the decoder with S-parameter labels in existing domains to provide physics-aware supervision, and (3) fine-tuning the foundation model with a replaced adapter for new and diverse tasks.

\vspace{-0.5em}
\subsection{Construction of Passive Design Datasets}
\label{sec:prelim:dataset_generation}

   \begin{table}[tb]
    \caption{Summary of Datasets Used in This Work}
    \centering
    \resizebox{1\linewidth}{!}{
        \begin{tabular}{ |c||c|c|c|c| }
        \hline
        & PulseRF~\cite{PulseRF} &   MOTIF-RF~\cite{MOTIF-RF}  & T-Coil &Coil\\   
        \hline
        \hline
        Tech Node &  InP &SOI&SiGe&GaN\\ 
        \hline
        Topology
        & Transformer & Transformer & Coupled Inductors & Spiral Coil \\
        \hline

        Key Features
        & Non-intuitive
        & Multi-template
        & Center-tapped
        & Standard \\
        \hline
            Frequency Range& 0--300\,GHz & 0--200\,GHz & 0--100\,GHz & 0--40\,GHz \\
            \hline
            
            Number of Ports&4&4&3&2\\
            \hline
            Simulation Tool&Ansys HFSS&Cadence EMX& Cadence EMX&ADS Momentum\\
            \hline
            Dataset Size &4k&30k&8k&9k\\
            \hline
            Layout Style& Pixelated &Vectorized&Vectorized& Pixelated\\
            \hline
        \end{tabular}
    }
    \label{tab:database}
    \vspace{-1.5em}
    \end{table}

\begin{figure}[t]
\centering
\includegraphics[
    width=1\linewidth,
    trim={10pt 10pt 10pt 10pt},
    clip
]{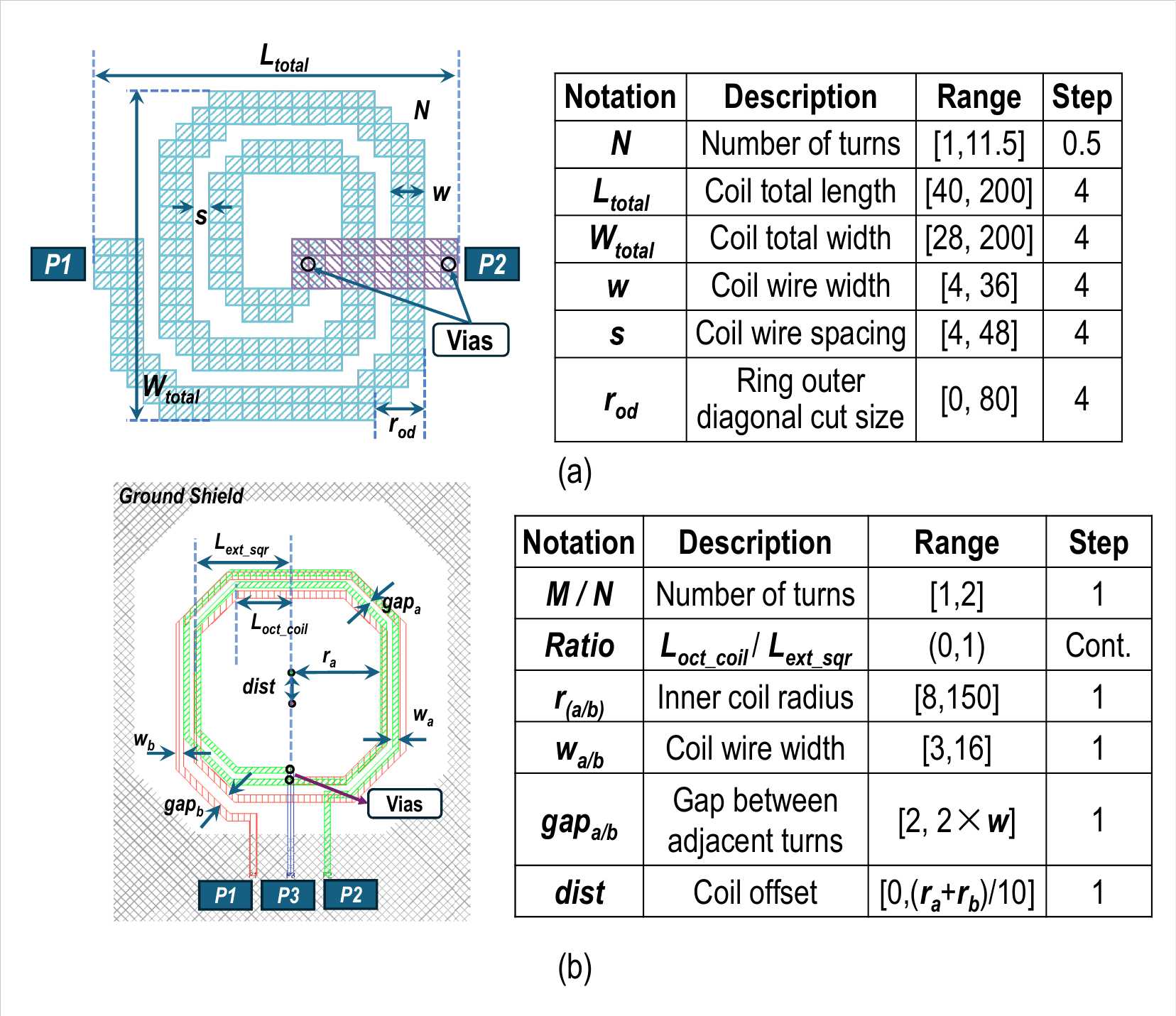}
\caption{(a) two-port inductors in Qorvo 150-nm GaN process and (b) three-port T-coils in GF 130-nm SiGe process.}
\label{fig:database}
\vspace{-1em}
\end{figure}


Generalization is the key objective of FoundaRF. Achieving this goal requires training and validation across diverse RF passive structures that vary widely in topology family, physical scale, semiconductor technology, number of ports, frequency, and complexity. 

Toward this end, this work includes four complementary datasets, as shown in Table \ref{tab:database}. These include (1) four-port transformer-seeded, non-intuitive structures in Teledyne 250-nm InP~\cite{PulseRF}, (2) four-port multi-template transformers in GF 22-nm CMOS SOI~\cite{MOTIF-RF},  and two newly created datasets: (3) three-port T-coils in GF 130-nm SiGe process, and (4) two-port inductors in Qorvo 150-nm GaN process. The two newly generated datasets are illustrated in Fig. \ref{fig:database}. 

Within each dataset, layout scale varies by up to 20$\times$, and the scale variation becomes even more significant across datasets.
Together, they provide the necessary diversity to thoroughly train and evaluate FoundaRF’s generalization capability.
Broadband S-parameters, simulated using industry-standard EM simulation software, are used as ground-truth labels for all datasets.

\begin{figure*}[t]
\centering
\includegraphics[trim=10 10 10 10, clip, width=0.95\textwidth]{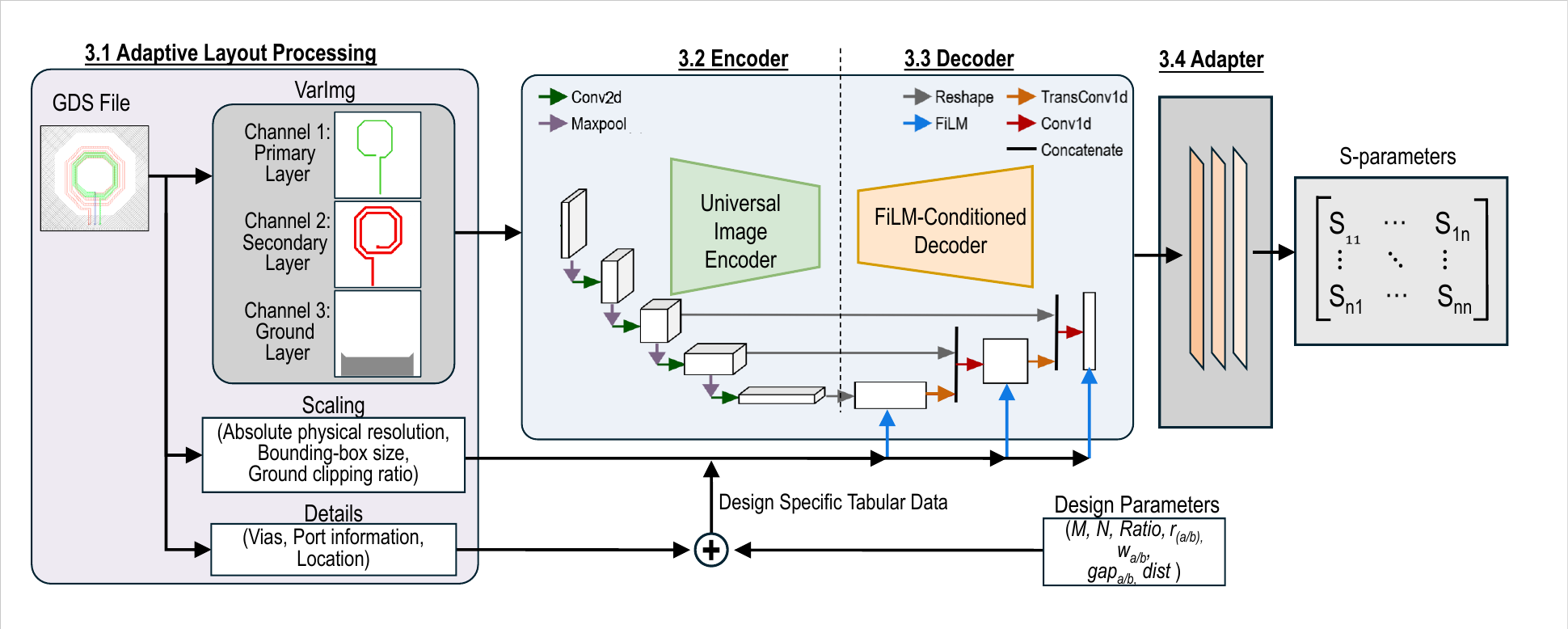}
\caption{FoundaRF overview. The framework consists of four stages: adaptive layout processing (Sec. \ref{sec:algo:overall}), a universal encoder (Sec. \ref{sec:algo:encoder}), a FiLM-conditioned decoder (Sec. \ref{subsec:decoder}), and a task-specific adapter that outputs the S-parameter matrix (Sec. \ref{subsec:adapter}).}
\label{fig:Model_Overview}
\vspace{-0.5em}
\end{figure*}

\section{FoundaRF Framework} \label{sec:algo}
This section presents the proposed FoundaRF framework. It consists of four stages: (1) a preprocessing stage that produces adaptive layout representations with scale-consistent features, (2) a universal convolutional encoder stage that is pretrained using a distance-field (DF) reconstruction objective, (3) a FiLM-conditioned decoder stage that integrates scale and design-specific data into encoded image, and (4) an adapter stage that enables task-specific projection to the S-parameter space. An overview is shown in Fig.~\ref{fig:Model_Overview}. The encoder and decoder structures inherit the UNet model in \cite{PulseRF}.
\vspace{-2em}


\vspace{-0.5em}
\subsection{Adaptive Layout Processing} \label{sec:algo:overall}
RF passive layouts vary significantly in physical dimensions and layout styles,
depending on performance targets, metal stacks, operating frequencies, and
designer preferences. Such diversity poses a fundamental challenge for
fixed-resolution rasterization, since small structures become under-resolved
while large layouts lose fine geometric detail. Prior work in computer vision
has shown that convolutional networks are highly sensitive to large variations in object scale: mismatched resolutions lead to aliasing, inconsistent feature
activation, and degraded representation quality
\cite{snip,scaleinv}. These studies consistently report that standard
CNNs struggle to learn scale-consistent features when inputs span a wide range
of physical sizes or when fine-grained patterns are compressed into too few pixels. 

To overcome this challenge, we propose the following adaptive layout processing pipeline. It begins with the original \textsc{GDSII} files, preserving the exact polygonal mask geometry regardless of whether the layout follows a pixelated or vectorized style. 
As illustrated in Fig.~\ref{fig:Model_Overview}, we extract three electromagnetically dominant metal layers from each layout: primary layer, secondary layer, and ground/shield plane, and rasterize them into a normalized three-channel image. 
Layout details that do not represent core topological geometry (e.g., via arrays and port locations) and scaling ratios (e.g., absolute physical resolution, bounding-box size, and ground clipping ratio) are stored separately as structured vectors and injected later at the decoder and adapter stage.

This adaptive layout processing pipeline offers three key advantages: (1)
\textbf{universal representation}, accommodating arbitrary RF passive
layouts, from compact mmWave inductors to large multi-turn transformers,
without suffering from fixed-resolution limitations; (2) \textbf{robust feature
extraction}, normalizing each layout to a comparable spatial scale preventing
the ``filter--scale mismatch’’ problem that plagues fixed-resolution CNNs \cite{snip,scaleinv} and thereby enabling the encoder to learn
topology-driven patterns rather than pixel-size artifacts; and (3)
\textbf{physics-aligned scaling}, separating absolute physical scale (decoder) from topology (encoder) is physically grounded in Maxwell’s equations: topology dominates EM field distributions, while absolute physical scale shifts operating frequency. By excluding absolute scale from topology and re-introducing it only at the decoder stage, the model naturally embeds Maxwell's scaling property by design.

Together, these elements form a physics-informed, scale-consistent representation that supports robust learning across diverse RF passive structures. This forms the basis for our further development of the FoundaRF framework and pretraining methodology.


\vspace{-0.5em}
\subsection{Universal Encoder with Distance-Field (DF)-Based Self-Supervised Pretraining} \label{sec:algo:encoder}

\begin{figure}[t]
    \centering
    \includegraphics[width=1\linewidth]{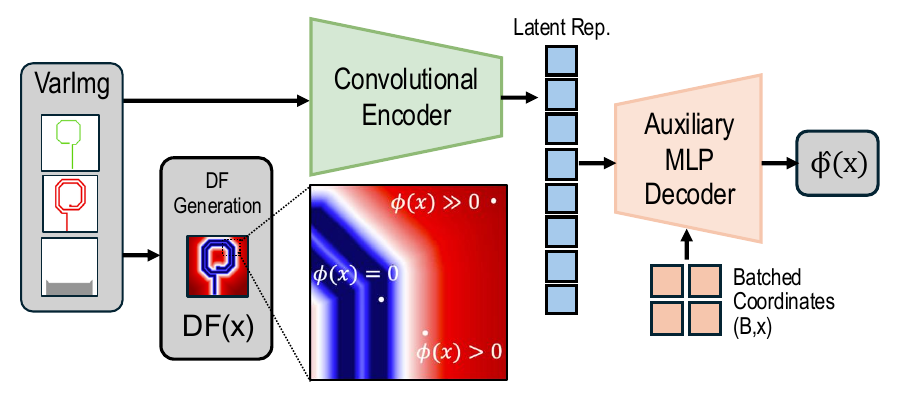}
    \caption{DF-Based Pretraining Overview.}
    \label{fig:DF_method}
\vspace{-1.5em}
\end{figure}

Full-wave EM simulation of a single RF passive structure requires minutes to hours depending on complexity, making large-scale labeled dataset generation prohibitively expensive. In contrast, unlabeled layout geometries are abundant across design libraries and process nodes. Self-supervised pretraining exploits this asymmetry: the encoder learns geometric representations from a large corpus of unlabeled layouts, \textbf{reducing the number of full solver simulations} required for downstream supervised fine-tuning.

A key observation is that, despite their wide diversity in topology and technology, all RF/mmWave passive structures are governed by the same Maxwell's equations, where EM field distributions, and the resulting S-parameters, are \textbf{dictated by proximity to conductor boundaries}. This makes the Distance Field (DF), the distance to the nearest conductor boundary at every spatial point, a natural training objective. DF can be computed from layout geometry alone without EM simulation labels, making it ideally suited for self-supervised training on large, diverse datasets of unlabeled layouts. During training, a lightweight decoder reconstructs DF from the encoder's latent representation, driving the encoder to learn spatial primitives (such as inter-trace proximity, conductor width, edge geometry) that govern EM field distributions across all passive structures. The resulting encoder provides a transferable geometric backbone, well-aligned with downstream S-parameter prediction regardless of topology, template, or process technology.

The Signed Distance Field (SDF) as a supervisory label has been previously 
adapted for shape representation in neural networks 
~\cite{Park_2019_CVPR}. Given a closed region $\Omega$ with boundary 
$\partial\Omega$, the SDF is defined as
\begin{equation}
    \psi(\mathbf{x}_0) = 
    \begin{cases}
        -\inf_{\mathbf{x}_s \in \partial\Omega} 
        \|\mathbf{x}_0 - \mathbf{x}_s\|_2  
            ,& \mathbf{x}_0 \in \Omega \\
        +\inf_{\mathbf{x}_s \in \partial\Omega} 
        \|\mathbf{x}_0 - \mathbf{x}_s\|_2 
            ,& \mathbf{x}_0 \notin \Omega \\
    \end{cases}
    \label{eq:sdf}
\end{equation}
where $\partial\Omega = \bigcup_{k} \partial\Omega_k$ is the union of all 
conductor boundaries in the layout, $\mathbf{x}_0 \in \mathbb{R}^2$ is an arbitrary query point, $\mathbf{x}_s \in \partial\Omega$ denotes a point on the conductor boundary, and the sign encodes whether a point lies inside or outside the boundary. 
However, unlike closed surfaces for which the SDF is formally defined, RF 
layouts contain many open boundaries, vias, and non-manifold junctions where 
a typical ``sign'' convention (inside vs.\ outside) is not well-defined. 
Applying a signed field under these circumstances introduces discontinuities 
and unstable gradients, detrimental for reconstruction-based self-supervision 
near edges. To address this, we discard the sign and define the unsigned 
distance field $\phi: \mathbb{R}^2 \to \mathbb{R}_{\geq 0}$ as
\begin{equation}
    \phi(\mathbf{x}_0) = \inf_{\mathbf{x}_s \in \partial\Omega} 
    \|\mathbf{x}_0 - \mathbf{x}_s\|_2
    \label{eq:udf}
\end{equation}
By removing the sign, $\phi$ is 
well-defined regardless of boundary topology and avoids the discontinuous 
sign transition at $\partial\Omega$ that makes the SDF ill-suited to open geometries. The field $\phi$ is Lipschitz-1 continuous, 
ensuring bounded gradients throughout the domain, and satisfies the Eikonal 
equation $\|\nabla\phi\| = 1$ a.e., providing a
numerically stable regression target for self-supervised pretraining. This 
yields a smooth geometry-derived signal while retaining key structural cues 
such as trace width, curvature, and proximity. 

An example DF generation is illustrated in Fig.~\ref{fig:DF_method}.
As a self-supervised learning task, DF Reconstruction has proven effective in downstream physics-driven surrogate modeling by leveraging spatial geometric features dependent on proximity~\cite{10.1115/DETC2025-168907}.
For the surrogate modeling of RF passives, features extracted from the DF align with differentiable geometric dependencies  (i.e., inter-trace distance) 
affecting EM field distributions.
This alignment is not merely geometric but is grounded in electromagnetic 
boundary physics. On perfect electric conductor (PEC) surfaces, the 
standard model for metallization in planar passive structures, the 
tangential electric field vanishes at the conductor boundary:
\begin{equation}
    \hat{n} \times \mathbf{E} = 0, \quad \mathbf{x} \in \partial\Omega
\vspace{-0.5em}
\end{equation}
Away from the boundary, the field magnitude recovers at a rate governed 
by distance to the nearest conductor surface. For a locally smooth boundary, 
the field grows linearly, $|\mathbf{E}| \propto \phi(\mathbf{x})$; near 
conducting edges, present at every trace corner and termination, the 
Meixner edge condition \cite{meixner1972} dictates a 
singular dependence $|\mathbf{E}| \propto \phi(\mathbf{x})^{\alpha}$, where 
$\alpha < 0$ is set by the local wedge geometry. At RF frequencies, field 
penetration into the conductor itself decays as 
$e^{-\phi(\mathbf{x})/\delta}$, where $\delta = \sqrt{2/(\omega\mu\sigma)}$ 
is the skin depth~\cite{pozar}. In each regime, the spatial dependence of 
the EM field is governed by distance to the nearest conductor boundary, 
precisely the quantity encoded by $\phi$. By reconstructing $\phi$ during 
pretraining, the encoder is biased toward a feature space whose dominant 
spatial modes align with those of relevant field distributions.

The pretraining flow is illustrated in Fig.~\ref{fig:DF_method}. We utilize a lightweight MLP decoder, as prior work has shown that a lightweight decoder is sufficient for the encoder to learn generalizing representations ~\cite{MIM_Wang}.
After pretraining, the encoder serves as a \textbf{geometric backbone agnostic to template and process variation}. Compared to conventional self-supervised learning methods with pixel-level reconstruction objectives, which encode rasterized intensity rather than metric geometry, DF reconstruction objectives encode features derived from smooth distance gradients. The variable image representation input is used as a standardized input to the universal encoder, which allows architectural consistency across templates and processes. As a result, the encoder is induced to reflect \emph{generalizable and differentiable geometric primitives} that appear consistently across templates.

\vspace{-0.5em}
\subsection{FiLM-Conditioned Decoder}
\label{subsec:decoder}

The latent representation extracted by the encoder in the previous subsection captures layout-dependent geometric features. However, this latent
representation is intentionally \emph{scale-invariant} and \emph{template-agnostic},
and therefore insufficient to directly predict physically accurate
S-parameters. 
To reintroduce the physical geometry
removed during adaptive layout normalization, we employ a \textbf{Feature-wise Linear Modulation
(FiLM)}-conditioned decoder. 

The decoder follows a hierarchical 1-D
convolutional architecture that mirrors the multi-scale structure of the
encoder, progressively reconstructing the broadband S-parameters from the
compressed latents.
At multiple stages within the decoder, FiLM blocks apply channel-wise affine
transformations of the form
\begin{equation}
    h' = (1 + \gamma)\odot h + \beta,
\end{equation}
where the modulation parameters $(\gamma, \beta)$ are produced by a small MLP
that processes a compact metadata vector. This vector encodes:
(1) the physical scaling ratio used during adaptive normalization; 
(2) template-specific design parameters; 
and (3) additional scalar layout details (e.g., via and port locations).

This scheme enables lightweight multimodal fusion by modulating visual features by non-visual tabular descriptors with minimal
overhead \cite{perez2018film}, allowing the decoder to model the projection between geometry-dependent behavior (captured by the encoder) and physics-dependent behavior (provided through modulation) in an expressive manner.


\subsection{Task-Specific Adapter Head}
\label{subsec:adapter}
While the decoder produces a rich, design-specific frequency-domain representation, different tasks require different output formats and dimensionalities. To support this without retraining the backbone, we introduce a lightweight \textbf{task-specific adapter head}.

The adapter consists of a compact sequence of 1-D convolutional \cite{STCNN} or linear layers
that project the decoder's shared representation into the final S-parameter
space. Depending on the prediction task, the adapter may produce outputs with dimensions subject to number of frequency points and number of ports.
Because the adapter is small in capacity and decoupled from the backbone, it can
be swapped or fine-tuned with minimal computational overhead.

This separation reinforces the modularity required for a foundation model.
The encoder and decoder together learn \emph{shared, technology-agnostic, and
physics-aware} representations, while the adapter captures the
\emph{task-specific projection} needed for a particular passive type, process node,
or output dimensionality. This modularity enables:
(1) rapid adaptation to new tasks by replacing only the adapter;
(2) low-cost fine-tuning since only a small number of parameters need to be
updated;
(3) stable multi-process and multi-template training; and
(4) consistent backbone behavior across diverse RF passive structures.




\vspace{-1em}
\section{FoundaRF Model Evaluation} \label{sec:results}

In this section, we present experimental results for~\workname~to demonstrate its effectiveness and reliability.  Model training costs around 2 minutes per 1000 samples on an NVIDIA L40 GPU. The UNet-based model adopted throughout \workname{} contains 5.2M parameters.
Experiments confirm that a 90M ViT-Base model achieves similar performance,
validating that the compact UNet backbone is sufficient for the dataset scales
considered in this work.
Sec.~\ref{sec:results:1} evaluates the proposed adaptive layout processing and the FiLM-conditioned decoder.
Sec.~\ref{sec:results:2} compares our proposed universal image encoder with baselines through few-shot experiments. 
Sec.~\ref{sec:results:3} presents the overall evaluation of \workname, demonstrating its cross-domain generalization.
The model performance is measured with the $R^2$ score and the mean absolute error (MAE) between the model prediction and EM-simulated ground truth, averaged across S-parameter dimensions in a similar way to \cite{MOTIF-RF}.

\vspace{-1em}
\subsection{Evaluation of Adaptive Layout Processing and FiLM-Conditioned Decoder} \label{sec:results:1}


\begin{table*}[t]
  \centering
  \caption{Model evaluation for adaptive layout processing and FiLM-conditioned
           decoder across datasets.}
  \label{tab:Model_by_input_representation}
  \resizebox{\linewidth}{!}{
  \begin{tabular}{|c|| c c | c c | c c |c  c |c c | c c | c c|}
  \hline
    \B Technology    &\multicolumn{8}{c|}{\B SOI}
                                & \multicolumn{2}{c|}{\B InP}
                                & \multicolumn{2}{c|}{\B SiGe}
                                & \multicolumn{2}{c|}{\B GaN}\\
    \hline
          \B Structure   & \multicolumn{2}{c}{\B 1:1}
                & \multicolumn{2}{c}{\B M:N}
                & \multicolumn{2}{c}{\B 8-shaped}
                & \multicolumn{2}{c|}{\B Parallel}
                & \multicolumn{2}{c|}{\B PulseRF}
                & \multicolumn{2}{c|}{\B T-Coil}
                & \multicolumn{2}{c|}{\B Coil}\\
    \hline
    \B Metric & MAE & $R^2$ & MAE & $R^2$ & MAE & $R^2$ & MAE & $R^2$ & MAE & $R^2$ & MAE & $R^2$ & MAE & $R^2$\\
    \hline
    \hline
    Tabular~\cite{MOTIF-RF}                     & 0.0214 & 0.978 & 0.0638 & 0.886 & 0.0108 & 0.992 & 0.0099 & 0.995 & 0.0302 & 0.919 & 0.0138 & 0.996 & 0.0041 & 0.997\\
     \hline
    Image~\cite{PulseRF}                       & 0.0193 & 0.982 & 0.0513 & 0.911 & 0.0125 & 0.990 & 0.0146 & 0.990 & 0.0250 & 0.921 & 0.0217 & 0.990 & 0.0074 & 0.997\\
    \hline
    Graph Transformer~\cite{MOTIF-RF}           & 0.0164 & 0.987 & 0.0614 & 0.890 & 0.0312 & 0.925 & 0.0583 & 0.871 & \multicolumn{2}{c|}{n/a$^\ddagger$} & 0.0139 & 0.995 & \multicolumn{2}{c|}{n/a$^\ddagger$}\\
    \hline
     Concat(VarImg, Scaling)    & 0.0089 & 0.996 & 0.0465 & 0.927 & 0.0072 & 0.996 & 0.0072 & 0.995 & 0.0183 & 0.950 & 0.0127 & 0.996 & 0.0028 & 0.999\\
    \hline
    FiLM(VarImg, Scaling)       & 0.0071 & 0.997 & 0.0439 & 0.927 & 0.0060 & 0.997 & 0.0059 & 0.998 & \textbf{0.0171} & \textbf{0.963} & 0.0102 & 0.997 & \textbf{0.0020} & \textbf{0.999}\\
    \hline
    \B FiLM(VarImg, Scaling)       &\multirow{2}{*}{\B0.0061}
                                & \multirow{2}{*}{\B0.997}
                                & \multirow{2}{*}{\B0.0406}
                                & \multirow{2}{*}{\B0.932}
                                & \multirow{2}{*}{\B0.0040}
                                & \multirow{2}{*}{\B0.999}
                                & \multirow{2}{*}{\B0.0030}
                                & \multirow{2}{*}{\B0.999}
                                & \multirow{2}{*}{\B0.0170}
                                & \multirow{2}{*}{\B0.955}
                                & \multirow{2}{*}{\B0.0078}
                                & \multirow{2}{*}{\B0.999}
                                & \multirow{2}{*}{\B0.0019}
                                & \multirow{2}{*}{\B0.999}\\
    \B +FiLM(VarImg, Tabular) &&&&&&&&&&&&&&\\
    \hline
    \B MAE improvement$^\dagger$
                & \multicolumn{2}{c|}{\B2.7$\times$}
                & \multicolumn{2}{c|}{\B1.3$\times$}
                & \multicolumn{2}{c|}{\B2.7$\times$}
                & \multicolumn{2}{c|}{\B3.3$\times$}
                & \multicolumn{2}{c|}{\B1.5$\times$}
                & \multicolumn{2}{c|}{\B1.8$\times$}
                & \multicolumn{2}{c|}{\B2.2$\times$}\\
    \hline
  \end{tabular}
  }
  {\raggedright\footnotesize$^\dagger$MAE improvement of the final method over the best
   single-modality baseline (Tabular~\cite{MOTIF-RF}, Image~\cite{PulseRF}, or Graph
   Transformer~\cite{MOTIF-RF}, whichever is lower) per dataset.\par
   $^\ddagger$The graph transformer requires a polyline abstraction of the metal
   traces. The InP and GaN layouts are pixelated structures for which no such graph
   abstraction exists.\par}
\vspace{-0.4em}
\end{table*}

\begin{table*}[t]
\centering
\caption{Ablation Study: In-Domain Few-Shot Evaluation of Encoder Self-Supervised Learning Objectives}
\label{tab:ssl-ablation-mae}
\large
\setlength{\tabcolsep}{5pt}
\renewcommand{\arraystretch}{1.0}
\begin{adjustbox}{max width=0.99\textwidth}
\begin{tabular}{|l l||cc|cc|cc|cc|cc|cc|cc|cc|}
\hline
\multicolumn{2}{|c||}{\textbf{Technology}} &
\multicolumn{4}{c|}{\textbf{SOI}} &
\multicolumn{4}{c|}{\textbf{InP}} &
\multicolumn{4}{c|}{\textbf{SiGe}} &
\multicolumn{4}{c|}{\textbf{GaN}} \\
\hline
\multicolumn{2}{|c||}{\textbf{Label \%}}&
\multicolumn{2}{c}{1\%} & \multicolumn{2}{c|}{5\%} &
\multicolumn{2}{c}{1\%} & \multicolumn{2}{c|}{5\%} &
\multicolumn{2}{c}{1\%} & \multicolumn{2}{c|}{5\%} &
\multicolumn{2}{c}{1\%} & \multicolumn{2}{c|}{5\%} \\
\cline{1-18}
\multicolumn{2}{|c||}{\textbf{Metric}}&
MAE & $R^2$ & MAE & $R^2$ &
MAE & $R^2$ & MAE & $R^2$ &
MAE & $R^2$ & MAE & $R^2$ &
MAE & $R^2$ & MAE & $R^2$ \\
\hline\hline
\multicolumn{2}{|l||}{Fully Supervised} &
0.0833 & 0.787 & 0.0347 & 0.953 &
0.0906 & 0.546 & 0.0547 & 0.808 &
0.1719 & 0.468 & 0.0766 & 0.885 &
0.0454 & 0.953 & 0.0141 & 0.990 \\
\hline
Vanilla Masked & Base &
0.0815 & 0.809 & 0.0435 & 0.933 &
0.0928 & 0.468 & 0.0624 & 0.734 &
0.1652 & 0.504 & 0.0813 & 0.864 &
0.0372 & 0.958 & 0.0168 & 0.986 \\
Reconstruction & Base (Frozen) &
0.0957 & 0.728 & 0.0946 & 0.733 &
0.1002 & 0.344 & 0.0879 & 0.506 &
0.1807 & 0.423 & 0.1491 & 0.604 &
0.1476 & 0.504 & 0.0584 & 0.900 \\
\cite{SimCLR} & Tiny &
0.0946 & 0.733 & 0.0414 & 0.940 &
0.0903 & 0.489 & 0.0536 & 0.796 &
0.1593 & 0.552 & 0.0779 & 0.867 &
0.0358 & 0.957 & 0.0140 & 0.987 \\
& Tiny (Frozen) &
0.1174 & 0.616 & 0.0701 & 0.841 &
0.1034 & 0.324 & 0.0865 & 0.523 &
0.1825 & 0.416 & 0.1531 & 0.565 &
0.0699 & 0.865 & 0.0367 & 0.958 \\
\hline
\textbf{Distance Field} & Base &
\textbf{0.0459} & \textbf{0.936} & \textbf{0.0259} & \textbf{0.974} &
\textbf{0.0775} & \textbf{0.636} & \textbf{0.0502} & \textbf{0.817} &
0.1462 & 0.606 & 0.0669 & 0.903 &
0.0321 & 0.966 & \textbf{0.0129} & \textbf{0.991} \\
\textbf{Reconstruction} & Base (Frozen) &
0.0486 & 0.929 & 0.0291 & 0.967 &
0.0821 & 0.529 & 0.0562 & 0.779 &
0.1449 & 0.611 & \textbf{0.0631} & \textbf{0.915} &
\textbf{0.0276} & \textbf{0.971} & 0.0129 & 0.990 \\
\textbf{(Ours)} & Tiny &
0.0526 & 0.915 & 0.0296 & 0.966 &
0.0789 & 0.625 & 0.0507 & 0.814 &
\textbf{0.1397} & \textbf{0.628} & 0.0662 & 0.903 &
0.0374 & 0.950 & 0.0145 & 0.985 \\
& Tiny (Frozen) &
0.0541 & 0.908 & 0.0340 & 0.954 &
0.0842 & 0.563 & 0.0504 & 0.816 &
0.1454 & 0.608 & 0.0658 & 0.906 &
0.0321 & 0.962 & 0.0155 & 0.985 \\
\hline
\multicolumn{2}{|l||}{\textbf{MAE improvement}$^\dagger$} &
\multicolumn{2}{c|}{\textbf{1.8$\times$}} &
\multicolumn{2}{c|}{\textbf{1.3$\times$}} &
\multicolumn{2}{c|}{\textbf{1.2$\times$}} &
\multicolumn{2}{c|}{\textbf{1.1$\times$}} &
\multicolumn{2}{c|}{\textbf{1.1$\times$}} &
\multicolumn{2}{c|}{\textbf{1.2$\times$}} &
\multicolumn{2}{c|}{\textbf{1.3$\times$}} &
\multicolumn{2}{c|}{\textbf{1.1$\times$}} \\
\hline
\end{tabular}
\end{adjustbox}

{\raggedright\footnotesize$^\dagger$MAE improvement of the best Distance Field Reconstruction variant over the best baseline (Fully Supervised or Vanilla Masked Reconstruction, whichever is lower) per column.\par}
\vspace{-0.2em}
\end{table*}

The selection of layout representation is critical to the performance and generalization capability of ML models for passive structure modeling.
Prior work mainly relies on tabular template features \cite{hua2020,PACOSYT}, layout images \cite{KaushikNC,PulseRF}, or graph encodings \cite{MOTIF-RF,dina_filter}.
This section evaluates whether the proposed adaptive layout representation in Sec.~\ref{sec:algo:overall} generalizes across layout styles, physical scales, and metal stacks, and whether combining tabular- and image-based inputs outperforms either representation alone.

Experiments are conducted across multiple datasets, with identical training conditions used to compare baselines and variations of the proposed approach.
Table~\ref{tab:Model_by_input_representation} summarizes the results.
The performance (MAE,~$R^2$) of the tabular, image, and graph-transformer baselines  varies across datasets, and none is universally optimal. The graph transformer \cite{MOTIF-RF} is strong on simple vectorized structures (e.g., 1:1 SOI) but degrades on complex windings and is inapplicable to the pixelated InP and GaN layouts.
The (VarImg, Scaling) configuration corresponds to the proposed adaptive layout representation, which consistently improves accuracy over all baselines.
For example, on the 1:1 SOI dataset, the MAE decreases from 0.0193 for the image-only baseline to 0.0089, showing a 54\% error reduction.

The proposed adaptive layout representation is further evaluated using two integration strategies: simple feature concatenation~(Concat) and FiLM-based conditioning.
FiLM consistently outperforms Concat. For instance, on the SiGe dataset, the MAE decreases from 0.0127 to 0.0102; and on the GaN dataset, it decreases from 0.0028 to 0.0020, demonstrating an additional 20\% to 30\% error reduction.

The last row of the table shows results by further combining tabular-based inputs with adaptive scaling and FiLM.
This configuration consistently achieves the best performance across all datasets and is adopted as our final input processing method.

\vspace{-1em}
\subsection{Evaluation of Image Encoder} \label{sec:results:2}
A key motivation to develop a robust self-supervised pretraining task is to provide the universal encoder with general geometric features relevant to the design space prior to application-specific fine-tuning. We evaluate the robustness of the DF Reconstruction objective through an ablation study under strict few-shot conditions in-domain across technologies, in order to validate the alignment of the extracted geometric features to S-parameter prediction.

Table~\ref{tab:ssl-ablation-mae} compares DF Reconstruction with masked reconstruction ~\cite{He_2022_CVPR} and a fully supervised baseline. First, for each SSL objective, a \textbf{base} encoder pretrained on 60\% of all unlabeled layouts and a \textbf{tiny} encoder pretrained on 10\% are evaluated. With the pretrained encoder, the models are then fine-tuned on a standard scheme under strict label constraints \cite{SimCLR}. The proportions of 1\% and 5\% of labels were selected to emulate label-constrained environments. The fine-tuned models are tested against frozen variants which isolate the learned representations. 
Across all four technologies and both label budgets, DF Reconstruction achieves the lowest MAE, outperforming both vanilla masked reconstruction and the fully supervised baseline in every few-shot setting. On the SOI dataset at the 1\% label budget, for example, it reduces the MAE from 0.0815 (vanilla masked reconstruction) to 0.0459, a 1.8$\times$ accuracy improvement.

The remaining rows of Table~3 explain where this advantage comes from. Frozen DF encoders retain most of their fine-tuned accuracy, reaching $R^2$ values up to 0.967 on SOI and 0.816 on InP without any supervised adaptation, whereas masked-reconstruction encoders degrade sharply once frozen. The two objectives also scale differently with unlabeled pretraining data. Because masked reconstruction rewards pixel-level intensity, additional pretraining largely specializes the encoder on rasterized texture that does not transfer to S-parameter prediction, and the \textbf{tiny} encoder often matches or exceeds \textbf{base}. DF Reconstruction shows the expected \textbf{base} $>$ \textbf{tiny} trend, as more pretraining yields more transferable geometric features, with SiGe as the sole exception given its large layout distribution shift. These results confirm that the DF objective genuinely captures the geometric regularities governing EM behavior, yielding a stronger pretrained representation regardless of label availability or pretraining data scale.
\vspace{-1em}
\subsection{Evaluation of FoundaRF Framework} \label{sec:results:3}

This subsection evaluates the capability of our proposed foundation model framework.
To explore this, we first combine the datasets of two technologies, SOI and InP, and pretrain the encoder using self-supervised learning. 
The in-domain few-shot downstream evaluation is presented in Table~\ref{tab:fewshot-22nm-inp}. When applying the pretrained encoder (trained on the combined dataset) to SOI and InP datasets individually, the results are as good as using encoders trained solely on their respective domains (as reported in Table 3). 
This validates that our model is able to accommodate diverse datasets across different semiconductor technologies and templates.

We further examine the model's ability to transfer across entirely new technologies and structures.
Fig.~\ref{fig:transfer_learning_dataset_size} presents results where the model, pretrained on combined SOI and InP datasets, is adapted to two unseen technologies, SiGe and GaN.
These target domains differ substantially in metal stack, layout characteristics, frequency range, and number of ports. 
Models trained from scratch consistently perform the worst under the same training sample size, while encoder-only transfer yields limited improvement, indicating that training solely on unlabeled images is insufficient to establish a meaningful warm start for S-parameter prediction on unseen structures, due to  the lack of physical supervision from decoder pretraining.

In contrast, transferring both the encoder and the decoder significantly improves performance, demonstrating that wideband S-parameter prediction benefits from supervised pretraining.
Moreover, by enabling fine-tuning through FiLM layers conditioned by new design-specific tabular data, the model achieves the best overall performance. This highlights the benefit of merging template-specific data with the template-invariant foundation backbone. 
FoundaRF demonstrates significant label-efficiency improvement. For instance, for GaN, comparable accuracy is achieved with just 25 samples, compared to 100 in the scratch setting, effectively reducing the data requirement by 4$\times$. The required samples scale with structural complexity, from a few dozen for the two-port inductors to a few hundred for the three-port T-coils.
These results confirm that our model can be efficiently adapted to new domains with minimal data, alleviating the need for extensive EM 
simulations.

\begin{table}[t]
\centering
\caption{Few-Shot Evaluation: (SOI+InP) Training Scheme}
\label{tab:fewshot-22nm-inp}
\scriptsize                          
\setlength{\tabcolsep}{3.75pt}        
\renewcommand{\arraystretch}{1.05}   

\begin{tabular}{|l||cc|cc|cc|cc|}
\hline
\multicolumn{1}{|c||}{\textbf{Technology}} &
\multicolumn{4}{c|}{\textbf{SOI}} &
\multicolumn{4}{c|}{\textbf{InP}} \\
\hline
\multicolumn{1}{|c||}{\textbf{Label Frac.}} &
\multicolumn{2}{c}{1\%} & \multicolumn{2}{c|}{5\%} &
\multicolumn{2}{c}{1\%} & \multicolumn{2}{c|}{5\%} \\
\hline
\multicolumn{1}{|c||}{\textbf{Metric}} &
MAE & $R^2$ & MAE & $R^2$ & MAE & $R^2$ & MAE & $R^2$ \\
\hline\hline

Fully Supervised &
0.0833 & 0.787 &
0.0347 & 0.953 &
0.0906 & 0.546 &
0.0547 & 0.808 \\
\hline
Frozen &
0.0519 & 0.917 &
0.0309 & 0.965 &
0.0898 & 0.496 &
0.0562 & 0.791 \\
\textbf{Fine-tuned} &
\textbf{0.0460} & \textbf{0.936} &
\textbf{0.0263} & \textbf{0.975} &
\textbf{0.0811} & \textbf{0.587} &
\textbf{0.0500} & \textbf{0.819} \\
\hline
\end{tabular}
\vspace{-2em}
\end{table}
\pgfplotstableread[row sep=\\,col sep=&]{
    Ns 	& Training from Scratch &Transfer(encoder only)& Transfer (encoder+TCN) & Transfer (encoder+ TCN) +template FiLM finetune\\
    25  & 0.2038 & 0.1972 & 0.1725 & 0.1460\\
    50  & 0.1595 & 0.1598 & 0.1334 & 0.1145\\
    100 & 0.1168 & 0.1132 & 0.0968 & 0.0836\\
    200 & 0.0875 & 0.0817 & 0.0670 & 0.0563\\
    400 & 0.0525 & 0.0578 & 0.0487 & 0.0375\\
    800 & 0.0364 & 0.0402 & 0.0336 & 0.0251\\
    1600 & 0.0221 & 0.0208 & 0.018 & 0.0142\\
    }\mydataTransferLearningSiGeMAE
\pgfplotstableread[row sep=\\,col sep=&]{
    Ns 	& Training from Scratch &Transfer(encoder only)& Transfer (encoder+TCN) & Transfer (encoder+ TCN) +template FiLM finetune\\
    25 & 0.2608 & 0.3177 & 0.458 & 0.6102 \\
    50 & 0.5398 & 0.5399 & 0.6599 & 0.7505 \\
    100 & 0.7414 & 0.7597 & 0.8134 & 0.8594 \\
    200 & 0.8464 & 0.8643 & 0.9116 & 0.9317 \\
    400 & 0.934 & 0.9239 & 0.9441 & 0.9674 \\
    800 & 0.9642 & 0.9608 & 0.972 & 0.9841 \\
    1600 & 0.9877 & 0.989 & 0.991 & 0.9949 \\
    }\mydataTransferLearningSiGeRtwo
\pgfplotstableread[row sep=\\,col sep=&]{
    Ns 	& Training from Scratch &Transfer(encoder only)& Transfer (encoder+TCN) & Transfer (encoder+ TCN) +template FiLM finetune\\
    25 &0.0747&0.0698&0.0607&0.0394\\
    50 &0.0545&0.0471&0.0400&0.0284\\
    100&0.0366&0.0378&0.0252&0.0202\\
    200&0.0224&0.0244&0.0160&0.0143\\
    400&0.0137&0.0148&0.0123&0.0103\\
    800&0.0101&0.0105&0.0082&0.0072\\
    }\mydataTransferLearningGaNMAE
\pgfplotstableread[row sep=\\,col sep=&]{
    Ns 	& Training from Scratch &Transfer(encoder only)& Transfer (encoder+TCN) & Transfer (encoder+ TCN) +template FiLM finetune\\
    25  & 0.8507 & 0.8665 & 0.9049 & 0.9588 \\
    50  & 0.9156 & 0.9330 & 0.9502 & 0.9728 \\
    100 & 0.9471 & 0.9400 & 0.9724 & 0.9817 \\
    200 & 0.9762 & 0.9681 & 0.9832 & 0.9858 \\
    400 & 0.9882 & 0.9861 & 0.9885 & 0.9893 \\
    800 & 0.9924 & 0.9909 & 0.9925 & 0.9926 \\
    }\mydataTransferLearningGaNRtwo

\newcommand{\figTransferLearningSiGeMAE}{
\begin{tikzpicture}
        \begin{axis}[
            width=0.52\linewidth,
            height=0.4\linewidth,
            scaled y ticks = false,
            yticklabel style={
                /pgf/number format/precision=2,
            },
            ylabel style={yshift= -3pt},
            ylabel=MAE,
            grid=both,
            xmode=log,              
            log basis x={10}       
        ]
            \addplot[dashed, color=red, mark=square*, mark options={fill opacity=0.3}] table[x=Ns,y=Training from Scratch ]{\mydataTransferLearningSiGeMAE};
            \addplot[densely dashed,color=cyan, mark=triangle*, mark options={fill opacity=0.3}] table[x=Ns,y=Transfer(encoder only) ]{\mydataTransferLearningSiGeMAE};
            \addplot[densely dashed, color=green!70!black, mark=diamond*, mark options={fill opacity=0.3}] table[x=Ns, y=Transfer (encoder+TCN) ]{\mydataTransferLearningSiGeMAE};
            \addplot[solid, color=blue, mark=*] table[x=Ns,y=Transfer (encoder+ TCN) +template FiLM finetune ]{\mydataTransferLearningSiGeMAE};

            
        \end{axis}
    \end{tikzpicture}
}

\newcommand{\figTransferLearningSiGeRtwo}{
\begin{tikzpicture}
        \begin{axis}[
            width=0.52\linewidth,
            height=0.4\linewidth,
            scaled y ticks = false,
            yticklabel style={
                /pgf/number format/precision=2,
            },
            ylabel=$R^2$,
            ylabel style={yshift= -3pt},
            grid=both,
            xmode=log,              
            log basis x={10}       
        ]
            \addplot[dashed, color=red, mark=square*, mark options={fill opacity=0.3}] table[x=Ns,y=Training from Scratch ]{\mydataTransferLearningSiGeRtwo};
            \addplot[densely dashed,color=cyan, mark=triangle*, mark options={fill opacity=0.3}] table[x=Ns,y=Transfer(encoder only) ]{\mydataTransferLearningSiGeRtwo};
            \addplot[densely dashed, color=green!70!black, mark=diamond*, mark options={fill opacity=0.3}] table[x=Ns, y=Transfer (encoder+TCN) ]{\mydataTransferLearningSiGeRtwo};
            \addplot[solid, color=blue, mark=*] table[x=Ns,y=Transfer (encoder+ TCN) +template FiLM finetune ]{\mydataTransferLearningSiGeRtwo};

            
        \end{axis}
    \end{tikzpicture}
}

\newcommand{\figTransferLearningGaNMAE}{
 \begin{tikzpicture}
        \begin{axis}[
            width=0.52\linewidth,
            height=0.4\linewidth,
            scaled y ticks = false,
            yticklabel style={/pgf/number format/fixed},
            ylabel=MAE,
            ylabel style={yshift= -3pt},
            grid=both,
            xmode=log,              
            log basis x={10}       
        ]
            \addplot[dashed, color=red, mark=square*, mark options={fill opacity=0.3}] table[x=Ns,y=Training from Scratch ]{\mydataTransferLearningGaNMAE};
            \addplot[densely dashed,color=cyan, mark=triangle*, mark options={fill opacity=0.3}] table[x=Ns,y=Transfer(encoder only) ]{\mydataTransferLearningGaNMAE};
            \addplot[densely dashed, color=green!70!black, mark=diamond*, mark options={fill opacity=0.3}] table[x=Ns, y=Transfer (encoder+TCN) ]{\mydataTransferLearningGaNMAE};
            \addplot[solid, color=blue, mark=*] table[x=Ns,y=Transfer (encoder+ TCN) +template FiLM finetune ]{\mydataTransferLearningGaNMAE};

            
        \end{axis}
    \end{tikzpicture}
}
\newcommand{\figTransferLearningGaNRtwo}{
 \begin{tikzpicture}
        \begin{axis}[
            width=0.52\linewidth,
            height=0.4\linewidth,
            scaled y ticks = false,
            yticklabel style={/pgf/number format/fixed},
            ylabel=$R^2$,
            ylabel style={yshift= -3pt},
            grid=both,
            xmode=log,              
            log basis x={10}       
        ]
            \addplot[dashed, color=red, mark=square*, mark options={fill opacity=0.3}] table[x=Ns,y=Training from Scratch ]{\mydataTransferLearningGaNRtwo};
            \addplot[densely dashed,color=cyan, mark=triangle*, mark options={fill opacity=0.3}] table[x=Ns,y=Transfer(encoder only) ]{\mydataTransferLearningGaNRtwo};
            \addplot[densely dashed, color=green!70!black, mark=diamond*, mark options={fill opacity=0.3}] table[x=Ns, y=Transfer (encoder+TCN) ]{\mydataTransferLearningGaNRtwo};
            \addplot[solid, color=blue, mark=*] table[x=Ns,y=Transfer (encoder+ TCN) +template FiLM finetune ]{\mydataTransferLearningGaNRtwo};

            
        \end{axis}
    \end{tikzpicture}
}
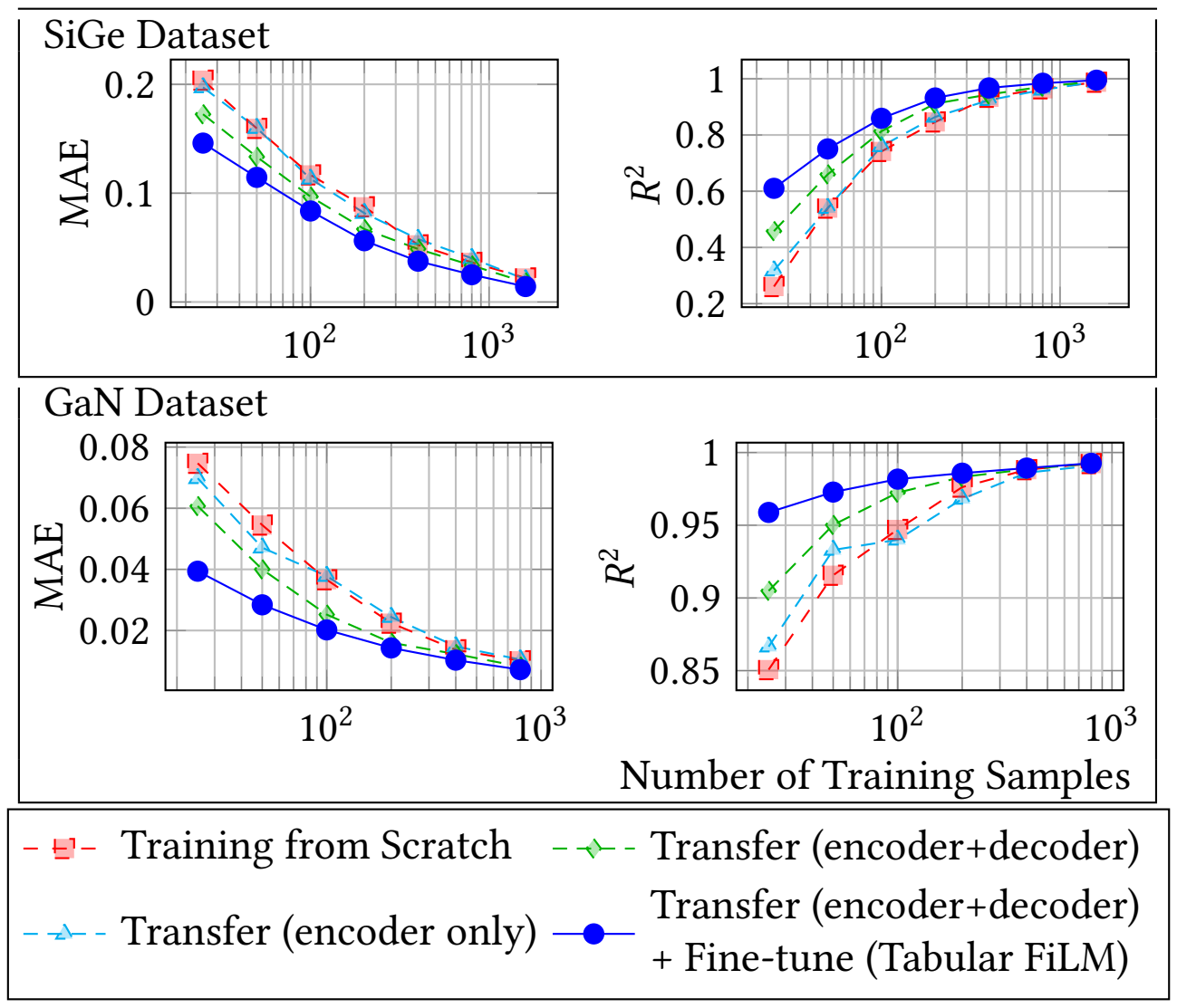
\begin{figure}[t]
\centering
{\renewcommand{\arraystretch}{0.2}
\begin{tabular}{|@{}c@{}@{}c@{}|}
\hline 
    \noalign{\vspace{2pt}}
   \multicolumn{2}{|l|}{SiGe Dataset}\\
    \figTransferLearningSiGeMAE  & \figTransferLearningSiGeRtwo \\
\hline
    \noalign{\vspace{2pt}}
    \multicolumn{2}{|l|}{GaN Dataset}\\
    \figTransferLearningGaNMAE   & \figTransferLearningGaNRtwo  \\
    \multicolumn{2}{|r|}{Number of Training Samples} \\ [2pt]
\hline
\end{tabular}    
}
\begin{tikzpicture}
\begin{axis}[
    hide axis,
    xmin=0, xmax=1,
    ymin=0, ymax=1,
    legend columns=2, 
    legend cell align=left,
    legend style={
        draw=black,
        /tikz/every node/.style={align=left},
        anchor=center,
    },
]

\addlegendimage{dashed, color=red, mark=square*, mark options={fill opacity=0.3}}
\addlegendentry{Training from Scratch}

\addlegendimage{densely dashed, color=green!70!black, mark=diamond*, mark options={fill opacity=0.3}}
\addlegendentry{Transfer (encoder+decoder)}

\addlegendimage{densely dashed, color=cyan, mark=triangle*, mark options={fill opacity=0.3}}
\addlegendentry{Transfer (encoder only)}

\addlegendimage{solid, color=blue, mark=*}
\addlegendentry{Transfer (encoder+decoder) \\+ Fine-tune (Tabular FiLM)}

\end{axis}
\end{tikzpicture}
\caption{Model performance vs. sample size for transfer learning from SOI+InP to other technologies (SiGe/GaN).}
\label{fig:transfer_learning_dataset_size}
\vspace{-1.5em}
\end{figure}

\vspace{-0.5em}
\section{FoundaRF Model Application}\label{sec:application}

While Sec.~\ref{sec:algo} and Sec.~\ref{sec:results} establish the
architecture and generalization capability of \workname{}, an equally important
question is what labels the model should learn and predict. In RF
passive modeling, this choice directly impacts model generality, data
efficiency, and physical consistency. In this section, we first justify the use
of broadband S-parameters as the universal prediction target for \workname{}
(Sec.~\ref{sec:design_spec}), and show both theoretically and empirically why
S-parameters are preferable to direct equivalent circuit parameter prediction.
We then demonstrate how this representation and the ~\workname{} model enable practical downstream
applications through two inverse design examples (Sec.~\ref{sec:inverse_}).

\vspace{-1em}
\subsection{S-parameters as a Universal Representation}
\label{sec:design_spec}

A fundamental design choice in \workname{} is to predict broadband S-parameters
rather than equivalent circuit parameters such as inductance or self-resonant
frequency directly, as in \cite{PACOSYT,yaolongIMS}. This choice is central to enabling generalization across
diverse RF passive structures, and we justify it from two complementary
perspectives: universality and learning efficiency through supervisory signal.

\begin{figure}[t]
    \centering
    \includegraphics[width=1\linewidth]{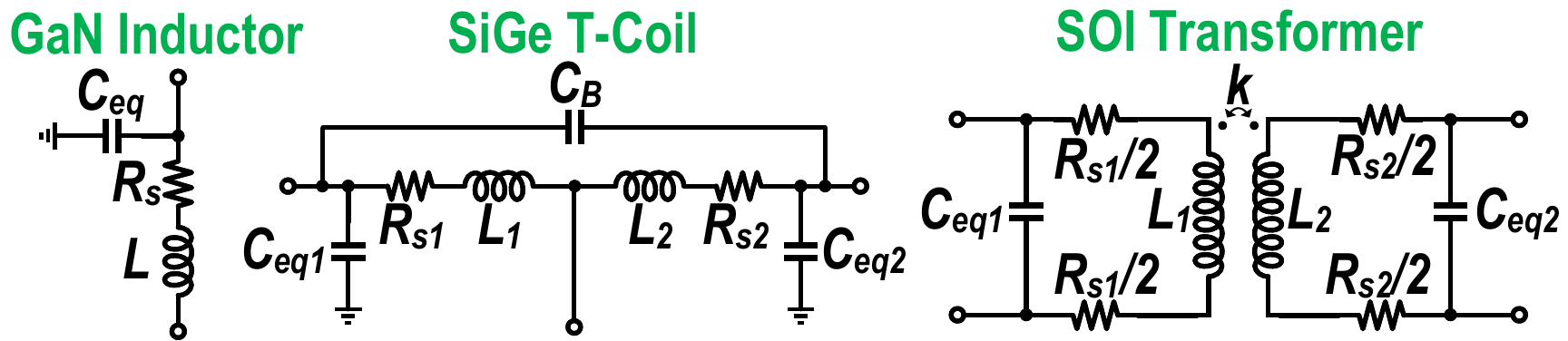}
    \caption{Schematic models for all three component families.}
    \label{fig:schematics}
    \vspace{-1em}
\end{figure}

\textbf{Universality and Physical Completeness.}
S-parameters provide a complete, broadband, port-wise electrical description of
any passive network. All standard equivalent circuit parameters---inductance
$L$, series resistance $R_s$, coupling coefficient $k$, self-resonant frequency
$f_\text{SRF}$, and bridge capacitance $C_B$---can be analytically derived from
S-parameters via S-to-Y conversion:
\begin{equation}
  \mathbf{Y} = \bigl(\mathbf{I} - \mathbf{S}\bigr)\bigl(\mathbf{I} +
  \mathbf{S}\bigr)^{-1} Y_0,
  \label{eq:s2y}
\vspace{-0.5em}
\end{equation}
where $Y_0 = 1/(50\,\Omega$), and circuit parameters are extracted as:
\begin{gather}
  L(f) = \frac{-1}{\omega \cdot \operatorname{Im}(Y_{ii})}, \qquad
  R_s(f) = \frac{\operatorname{Re}(Y_{ii})}{|Y_{ii}|^2}, \label{eq:LRs} \\
  f_{\mathrm{SRF}} = \min\bigl\{\, f : \operatorname{Im}\bigl(Y_{ii}(f)\bigr) = 0 \,\bigr\} \label{eq:SRF}
\end{gather}
\vspace{-0.5em}
and the coupling coefficient between ports $i$ and $j$ as:
\begin{equation}
  k_{ij} = \frac{-\operatorname{Im}(Y_{ij})}
           {\sqrt{\operatorname{Im}(Y_{ii}) \cdot \operatorname{Im}(Y_{jj})}}.
  \label{eq:k}
\end{equation}
Note that quality factor $Q = \omega L / R_s$ and parasitic capacitance
$C_\text{eq} = 1/(4\pi^2 f_\text{SRF}^2 L_\text{DC})$ both follow directly
from the above and are not independent parameters. The schematic circuit models for all three component families are illustrated in
Fig.~\ref{fig:schematics}.

This establishes a one-way relationship: S-parameters $\rightarrow$ equivalent
circuit parameters, but not the reverse. Equivalent circuit models are valid
only below a technology- and topology-dependent frequency limit ($f_\text{SRF}$), beyond which distributed coupling, parasitic
resonances, and high-frequency losses invalidate the lumped abstraction.
Moreover, certain passive structures, such as the non-intuitive InP layouts in~\cite{PulseRF}, do not admit a well-defined analytical equivalent circuit
model at all, and are excluded from equivalent circuit parameter evaluation for
this reason. S-parameters, by contrast, encode the complete frequency-domain
behavior across the full simulation bandwidth, making them the only
representation that generalizes across all topologies and frequency ranges
in \workname{}.


\textbf{Supervisory Signal and Learning Efficiency.}
Beyond physical completeness, S-parameters provide a richer and more consistent
supervisory signal for learning. Equivalent circuit parameters such as $L(f)$
and $R_s(f)$ are frequency-dependent curves valid only at low frequency,
while scalar metrics such as $f_\text{SRF}$ and $C_B$ can only be extracted
from a subset of designs where the lumped abstraction holds. In contrast,
S-parameters are defined over the full simulation bandwidth for every design,
providing dense, physics-grounded supervision at every frequency point. For the
GaN 2-port dataset (0--40\,GHz at a 0.25-GHz step), each sample contributes
$160 \times 4 = 640$ S-parameter values, compared to a much smaller set of
equivalent circuit targets that are only valid at low frequency. 

S-parameter prediction also aligns naturally with the foundation model objective
of \workname{}. As shown in Sec.~\ref{sec:results:3}, transferring
both the encoder and S-parameter decoder to unseen domains significantly
outperforms encoder-only transfer, confirming that supervised S-parameter
pretraining accelerates cross-domain adaptation. Their smoothness, boundedness,
reciprocity, and passivity make them stable regression targets that implicitly
encode physical constraints. Crucially, S-parameters provide a unified output
representation across all passive families: only the lightweight adapter head
needs to be replaced when adapting to a new technology, accommodating different
frequency ranges and port counts without retraining the backbone.

To quantify these advantages empirically, we compare two strategies using the
same \workname{} architecture and training budget: \textbf{S-param prediction},
which predicts broadband S-parameters and derives all equivalent circuit
parameters analytically via Eqs.~(\ref{eq:s2y})--(\ref{eq:k}), and
\textbf{direct prediction}, which predicts equivalent circuit parameters as
the model output directly with labels derived from ground-truth S-parameters
using the identical formulas. In the direct
prediction approach, only the adapter head is replaced.

Table~\ref{tab:spec_compare} reports MedPE and per-sample $R^2$ for both
strategies across SOI 1:1, SiGe, and GaN. All evaluations are performed at $f < 0.7 \times f_\text{SRF}$. Symmetric metrics are averaged across
windings or branches. S-param prediction achieves sub-percent MedPE on $L$ and
$f_\text{SRF}$ across all three technologies, and sub-percent MedPE on $k$ for
the SOI transformer. The $R^2$ contrast is particularly revealing: SOI $L$
drops from 0.999 to 0.187, and SiGe $k$ collapses to 0.038, indicating the
direct predictor has essentially no discriminative power on these quantities by regressing toward the dataset mean.

\begin{table}[t]
  \centering
  \caption{S-param prediction (ours) vs.\ direct equivalent circuit parameter
           prediction. MedPE = median percentage error.}
  \label{tab:spec_compare}
  \resizebox{\linewidth}{!}{%
  \begin{tabular}{ll
                  S[table-format=2.2,table-space-text-post=\%]
                  S[table-format=1.3]
                  S[table-format=2.2,table-space-text-post=\%]
                  S[table-format=1.3]
                  S[table-format=2.1,table-space-text-post={$\times$}]}
    \toprule
    \multirow{2}{*}{\shortstack[l]{\textbf{Metric}\\[-1pt]{\footnotesize($^\ast$: average)}}}
    & & \multicolumn{2}{c}{\textbf{Direct pred.}} &
        \multicolumn{2}{c}{\textbf{S-param pred.\ (ours)}} & \\
    \cmidrule(lr){3-4}\cmidrule(lr){5-6}
    & {\textbf{Tech}} &
      {MedPE} & {$R^2$} & {MedPE} & {$R^2$} & {Improvement} \\
    \midrule
    \multirow{3}{*}{$L^\ast$}
      & SOI  &  6.39\% & 0.187 & 0.45\% & 0.999 & 14.2$\times$ \\
      & SiGe &  5.33\% & 0.856 & 0.56\% & 0.996 &  9.5$\times$ \\
      & GaN  &  2.09\% & 0.978 & 0.35\% & 0.999 &  6.0$\times$ \\
    \midrule
    \multirow{3}{*}{$R_s^\ast$}
      & SOI  & 23.53\% & 0.256 & 3.03\% & 0.995 &  7.8$\times$ \\
      & SiGe & 28.20\% & 0.626 & 7.83\% & 0.912 &  3.6$\times$ \\
      & GaN  &  5.44\% & 0.974 & 2.04\% & 0.996 &  2.7$\times$ \\
    \midrule
    \multirow{3}{*}{$f_\text{SRF}^\ast$}
      & SOI  &  1.01\% & 0.755 & 0.30\% & 0.999 &  3.4$\times$ \\
      & SiGe &  1.84\% & 0.725 & 0.27\% & 0.976 &  6.8$\times$ \\
      & GaN  & 28.32\% & 0.770 & 0.29\% & 0.999 & 97.7$\times$ \\
    \midrule
    \multirow{2}{*}{$k$}
      & SOI  &  5.89\% & 0.676 & 0.83\% & 0.998 &  7.1$\times$ \\
      & SiGe & 70.99\% & 0.038 & 5.01\% & 0.949 & 14.2$\times$ \\
    \midrule
    $C_B$ & SiGe & 18.93\% & 0.451 & 4.87\% & 0.921 &  3.9$\times$ \\
    \bottomrule
  \end{tabular}}
  \vspace{-1em}
\end{table}

\vspace{-1em}
\subsection{Inverse Design Example}
\label{sec:inverse_}

\begin{figure}[t]
    \centering
    \includegraphics[
        width=1\linewidth,
        trim=0.5cm 0.2cm 0.5cm 0.2cm,
        clip
    ]{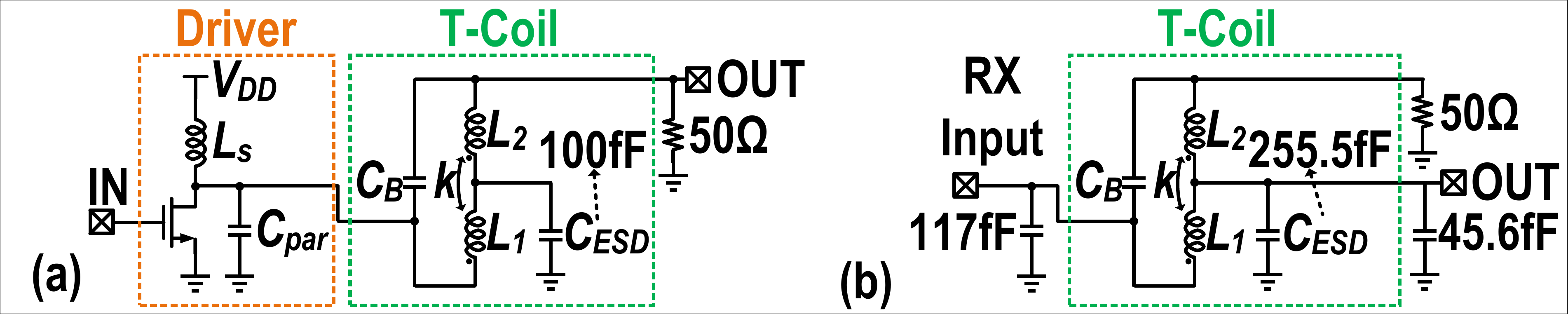}
    \caption{T-coils inverse design for (a) driver and (b) RX input.}
    \label{fig:Inverse_Design_schematic}
    \vspace{-1.3em}
\end{figure}
\begin{figure}[t]
    \centering
    \includegraphics[
        width=0.95\linewidth,
        trim=0.1cm 6.5cm 0.2cm 0.3cm,
        clip
    ]{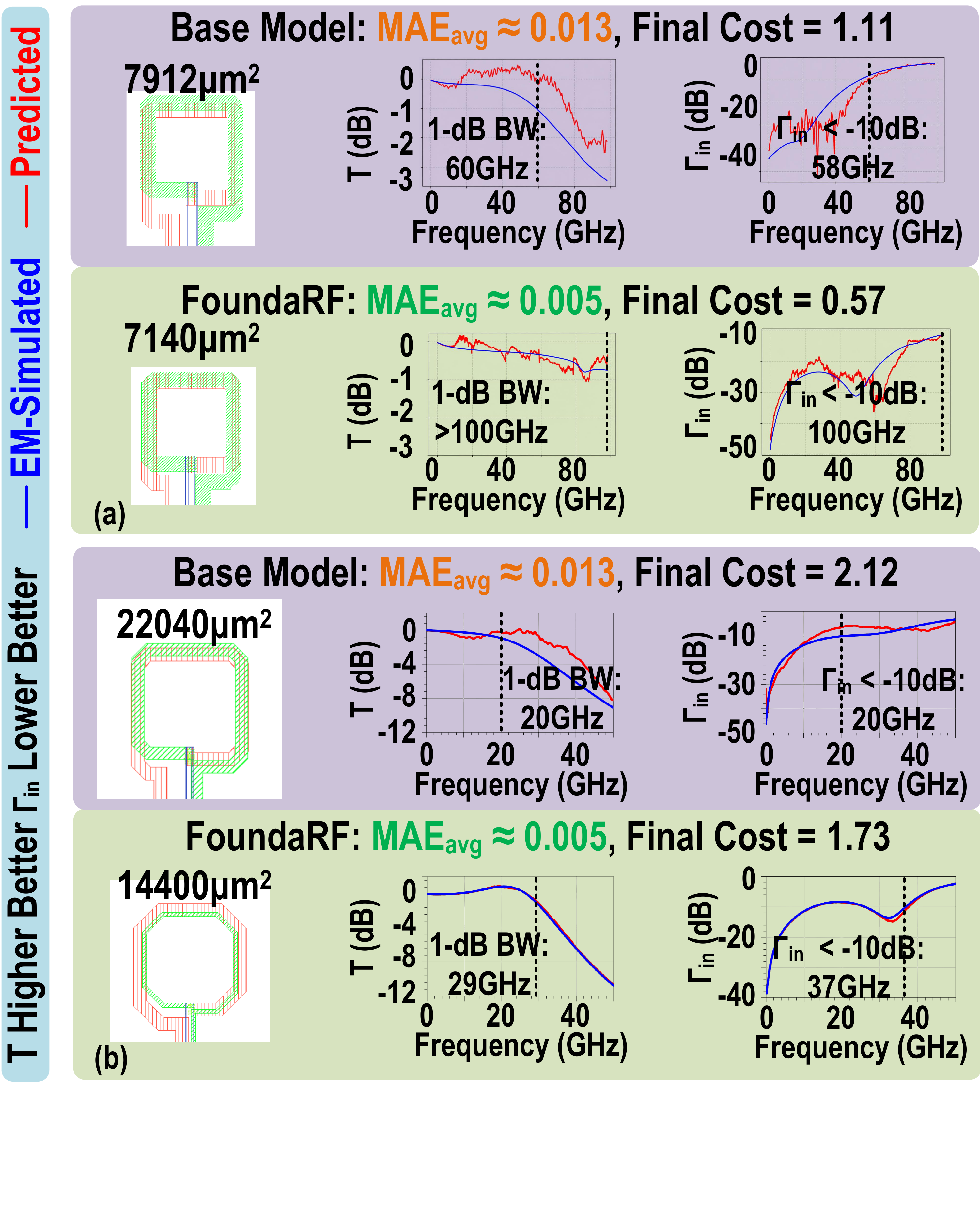}
    \caption{T-coil inverse design results for (a) driver and (b) RX input.}
    \label{fig:Inverse_Design}
    \vspace{-1.2em}
\end{figure}

A major application of ML surrogate models for passive components is inverse
design~\cite{KaushikNC}, i.e., automatically synthesizing GDSII from circuit
performance targets. To demonstrate the practicality of \workname{}, we apply
it to two real-world inverse design tasks: a driver output network and an RX
input ESD protection network, both using three-port T-coils as illustrated in
Fig.~\ref{fig:Inverse_Design_schematic}. T-coils are widely used in high-speed
wireline transceivers to absorb the parasitic capacitance of the ESD diode
connected to the center tap, thereby extending the transceiver
bandwidth~\cite{razavi_tcoil,tcoil_sudip}.

The optimization minimizes the cost function:
\begin{equation}\label{cost_function}
  J_S = w_0 A + \sum\nolimits_{f \in \Omega} w_f \Bigl(
        w_1 |\Gamma_\text{in}| + w_2 \bigl(1 - |T|\bigr)
        \Bigr),
\end{equation}
where $w_0$, $w_1$, $w_2$ are weights assigned to layout area $A$, input
reflection coefficient $\Gamma_\text{in}$, and transmission $T$, respectively,
and $w_f$ is a frequency-domain window function~\cite{MOTIF-RF}. Both
$\Gamma_\text{in}$ and $T$ are derived from the predicted S-parameters via
standard port-termination and network reduction, connecting the optimization
objective directly to the universal representation of Sec.~\ref{sec:design_spec}.
A well-designed circuit targets $|\Gamma_\text{in}| < {-}10\,\text{dB}$ and
$|T|$ as close to $0\,\text{dB}$ as possible across the band. We adopt the
Covariance Matrix Adaptation Evolution Strategy (CMA-ES)~\cite{cmaes}, which
dynamically adjusts both the covariance matrix and step size during
optimization. Using 4 CPU cores, \workname{} consistently converges to
near-optimal cost values within ten minutes, compared to hours or even days that would be required for EM-in-the-loop optimization.

Figure~\ref{fig:Inverse_Design} compares inverse design quality for different
modeling approaches trained on the same dataset. For the driver task
[Fig.~\ref{fig:Inverse_Design}(a)], \workname{} achieves a final cost of 0.57
and a layout area of 7{,}140\,$\mu$m$^2$, compared to a cost of 1.11 and
7{,}912\,$\mu$m$^2$ for the baseline model. For the RX input task
[Fig.~\ref{fig:Inverse_Design}(b)], \workname{} achieves a final cost of 1.73
and a layout area of 14{,}400\,$\mu$m$^2$, compared to 2.12 and
22{,}040\,$\mu$m$^2$ for the baseline. In both cases, \workname{} predictions
closely match EM-simulated ground truth, while the baseline model fails to
deliver reliable predictions. This result highlights that surrogate accuracy
impacts not only prediction error but also inverse-design fidelity:
under the same CMA-ES budget, \workname{} guides the search to an EM-verified
driver design exceeding 100\,GHz 1-dB bandwidth, whereas the baseline
converges to a substantially inferior local optimum at only 60\,GHz, a
40\% shortfall. The two tasks differ in circuit topology and load conditions
yet are handled by the same \workname{} model, demonstrating cross-task
generalization. This example further demonstrates that the S-parameter universality established in Sec.~\ref{sec:design_spec} translates directly into design productivity:
reliable broadband S-parameter predictions enable the optimizer to faithfully
navigate the design space, yielding verified high-performance layouts in
minutes without a single EM simulation during optimization.

\vspace{-0.5em}
\section{Conclusion} \label{sec:conclu}

This paper presents FoundaRF, a foundation-model framework for RF/mmWave passive component modeling that generalizes across diverse layout structures, physical scales, component families, and semiconductor technologies. FoundaRF processes raw GDSII layouts through an adaptive representation pipeline that supports scale-invariant and template-agnostic feature learning, pairing a universal DF-based encoder and a FiLM-conditioned decoder with lightweight, task-specific adapters to predict broadband S-parameters. Across four semiconductor technologies, it shows robust cross-technology and cross-structure transferability, including effective few-shot adaptation to unseen domains, and consistently outperforms all evaluated baselines.

We present FoundaRF as a framework rather than a claim that its compact 5.2M-parameter backbone is itself a large-scale foundation model. At present data scales, cross-domain transfer is driven by physics-aligned S-parameter supervision rather than model size, and the framework's value lies in its generalization and label efficiency. Several extensions follow naturally: scaling capacity with larger and more diverse pretraining data to approach the large-scale foundation-model regime, broadening beyond the magnetically-coupled families studied here to other EM structures such as transmission-line networks and couplers, and extending the 2.5D three-layer representation toward full 3D metal-stack modeling. 
Together, these establish FoundaRF as a scalable, high-fidelity backbone for RF/mmWave passive design automation.

\begin{acks}
The authors thank Xin Zhang, Luyao Shi, Prashanth Vijayaraghavan, and Ehsan Degan of IBM Research for their insightful discussions and feedback throughout this work.
\end{acks}


\clearpage
\balance

\bibliographystyle{unsrt}
\bibliography{./ref/TOP, ./ref/general, ./ref/aspdac26_ref, ./ref/rf_eda, ./ref/ML, ./ref/Adaptive_img.bib, ./ref/FiLM.bib}

\end{document}